\documentstyle[12pt,aasms4]{article}

\slugcomment{}

\lefthead{Chanam\'e et al.}  \righthead{The ionized gas kinematics of
NGC 1427A}

\begin{document}

\title{The Ionized Gas Kinematics of the LMC-type galaxy NGC 1427A in
the Fornax Cluster\footnote{Based on data collected at Las Campanas
Observatory, Chile, run by the Carnegie Institution of Washington}}

\author{J. Chanam\'e, L. Infante, and A. Reisenegger}
\affil{Departamento de Astronom\'\i a y Astrof\'\i sica, Pontificia
Universidad Cat\'olica de Chile, Casilla 306, Santiago 22, CHILE}

\begin{abstract}

NGC 1427A is a LMC-like irregular galaxy in the Fornax cluster with an
extended pattern of strong star formation around one of its edges,
which is probably due to some kind of interaction with the cluster
environment.  We present H$\alpha$ velocities within NGC 1427A,
obtained through long-slit spectroscopy at seven different positions,
chosen to fall on the brightest H\,{\sc ii} regions of the galaxy.
Due to its location very near the center of the cluster this object is
an excellent candidate to study the effects that the cluster
environment has on gas-rich galaxies embedded in it.  The rotation of
NGC 1427A is modeled in two different ways.  The global ionized gas
kinematics is reasonably well described by solid-body rotation,
although on small scales it shows a chaotic behaviour.  In this simple
model, the collision with a smaller member of the cluster as being
responsible for the peculiar morphology of NGC 1427A is very unlikely,
since the only candidate intruder falls smoothly into the general
velocity pattern of the main galaxy.  In a more elaborate model, for
which we obtain a better solution, this object does not lie in the
same plane of NGC 1427A, in which case we identify it as a satellite
bound to the galaxy.  These results are discussed in the context of a
normal irregular versus one interacting with some external agent.
Based on several arguments and quantitative estimates, we argue that
the passage through the hot intracluster gas of the Fornax cluster is
a very likely scenario to explain the morphological properties of NGC
1427A, although our kinematical data are not enough to support it more
firmly nor rule out the possibility of a normal irregular.

\end{abstract}


\section{Introduction}

Interactions between galaxies and their environments are thought to be
important mechanisms driving galaxy evolution.  For example, they have
been invoked to explain the excess of blue galaxies in high redshift
clusters relative to present-day clusters, the so-called
Butcher-Oemler effect (\cite{but78}; \cite{gun89}; \cite{evr91}).
Clusters of galaxies are ideal places to study these interactions, due
to their great concentration of galaxies of various morphologies,
sizes and luminosities, and huge masses of gas, in a comparatively
small volume of space.  Among the various kinds of interactions that
could be experienced by a cluster galaxy we have: tidal forces from
another galaxy or from the cluster as a whole (\cite{byr90};
\cite{hen96}), the ram pressure from the passage through the
intracluster medium (ICM) (\cite{gun72}; \cite{gio85}; \cite{evr91};
\cite{pho95}), high-speed encounters between galaxies (\cite{moo96}),
collisions and mergers (\cite{lyn76}; \cite{the77}; \cite{bar91}), and
the combined action of two or more of these mechanisms (\cite{pat92};
Lee, Kim, \& Geisler 1997).

The Fornax cluster is a relatively poor galaxy cluster dominated by
early-type galaxies.  Compared to Virgo, the center of Fornax is two
times denser in number of galaxies, but Virgo as a whole is almost
four times richer (\cite{fer88}; \cite{hil98}).  The hot ICM of Fornax
shines in X-rays, as detected by ROSAT and ASCA (\cite{jon97};
\cite{ran95}; \cite{ike96}), and this hot gas extends at least 200 kpc
from the center of the cluster.  Two giant ellipticals, NGC 1399 (a cD
galaxy with an extended halo of about 400 kpc in diameter, and an
extraordinarily large population of globular clusters, see
\cite{hil98}; \cite{gri99}) and NGC 1404, lie at the center of the
cluster.  Fornax may be composed of two subclusters in the process of
merging, evidenced by the big relative radial velocity between NGC
1399 and NGC 1404 of about 500 km/s (Bureau, Mould, \& Staveley-Smith
1996).  However, these galaxies are close in space.  Distance
determinations based on surface brightness fluctuations (Jensen,
Tonry, \& Luppino 1998) and globular cluster luminosity functions
(\cite{ric92}; \cite{gri99}) put them at roughly the same distance.
Moreover, the X-ray observations with ROSAT show that the hot corona
associated with NGC 1404 is distorted and probably being stripped,
indicating an infall of this galaxy towards NGC 1399 and the cluster
center (\cite{jon97}).

NGC 1427A is the brightest irregular (Irr) galaxy in the Fornax
cluster, and very similar to the LMC in its morphology and colors
(\cite{hil97}).  The great majority of the high surface brightness
regions that dominate the light of NGC 1427A are aligned along the
south-western edge of the galaxy, in a kind of distorted ring (see
Fig.1 and Fig.5).  Several arguments point towards explaining the
appearance of this galaxy in the context of an interaction with its
environment.  The resemblance to the so-called ring galaxies led
Cellone and Forte (1997) to suggest that NGC 1427A is the result of an
encounter with a smaller intruder, giving also a candidate for this
intruder (the North Object, see Fig. 1 and Fig. 5).  NGC 1427A is also
very close to the center of the cluster, with a projected distance of
121 kpc to NGC 1399 and 83 kpc to NGC 1404 \footnote{Throughout this
paper we assume a distance to the Fornax cluster of 18.2 Mpc, from
Kohle et al. (1996), recalibrated as in Della Valle et al. (1998)
using the new distances to Galactic globular clusters from Hipparcos
(\cite{gra97}).}, so tidal forces might be important in the
enhancement of the star formation in the galaxy.  Finally, NGC 1427A
is crossing the ICM of Fornax at a supersonic speed (see Section 4),
so the ram pressure exerted by the intracluster gas could also be the
cause of the peculiar distribution of star forming regions in the
galaxy.  Gavazzi et al. (1995) studied three galaxies in the cluster
Abell 1367 which, like NGC 1427A, have their bright H\,{\sc ii}
regions distributed along one edge of their perimeters, which they
attribute to the increase of the external pressure as the galaxies
cross the ICM.

In this paper we present the kinematics of the ionized gas (H\,{\sc
ii} regions) of NGC 1427A and discuss the obtained velocity field in
the context of a normal Irr galaxy versus an interacting galaxy.  In
Section 2 we describe the observations, reduction of the data and the
error analysis.  In Section 3 we model the kinematics of the galaxy
and analize the results.  Section 4 contains the discussion of the
possible scenarios for the history of NGC 1427A in the light of our
results, and in Section 5 we give our conclusions.

\section{Observations, Data Reduction and Error Analysis}

Long-slit spectra of NGC 1427A were obtained during two runs with the
2.5m DuPont telescope at Las Campanas Observatory, Chile, during 1997
February 3-4 and August 9-14.  The telescope was equipped with the
Modular Spectrograph.  The grating used had 600 grooves/mm, and as the
detector we used a 2048$\times$2048 SITe chip, with a pixel size of 15
$\mu$m.  This setup gives a dispersion of 1.27 $\AA$/pix and a spatial
sampling of 0.3625 arcsec/pix.  On the February run the measured
seeing was about 1 arcsec during the entire night, which corresponds
to a linear scale of 88 parsec at the adopted distance to Fornax.  For
the August run, due to the presence of some clouds, we binned the
spatial direction by a factor of 2 in order to get more light,
obtaining 0.725 arcsec/pix.  The seeing was 1.4 arcsec, resulting in a
spatial resolution of 123 parsec.  Integration times were of 45
minutes at the slit positions where three 15-minute frames were
obtained, and 15 minutes otherwise.  The instrumental resolution was
derived by measuring the FWHM of several unblended lamp lines after
calibrations.  For the February run we obtained a mean FWHM of 2.98
$\AA$, corresponding to a standard deviation of the Gaussian $\sigma$
= 1.27 $\AA$ (i.e., 58 km/s at H$\alpha$), and a mean FWHM of 4.8
$\AA$ for the August run, corresponding to $\sigma$ = 2.05 $\AA$
(i.e., 93 km/s at H$\alpha$).  The wavelength range is 4700 $\AA$ -
6850 $\AA$ for the February run and 4800 $\AA$ - 6960 $\AA$ for the
August run.  This range includes several emission lines of the ionized
gas, namely, H$\beta$, [OIII], HeI, [NII], H$\alpha$, and [SII] (see
Fig. 2).  The slit was aligned in order to cover the majority of the
bright H\,{\sc ii} regions of the galaxy.  The positions of the slits
are shown in Fig. 1 and were derived by matching coordinate
information obtained on the guider screen during the observations with
an H$\alpha$ image of the galaxy.  The coincidence between the spatial
profiles along the slits and their inferred positions on the galaxy
was almost perfect.  The images show the strong emission lines of the
H\,{\sc ii} regions, but very weak emission coming from the regions
between them, so we are mostly restricted to work with the brightest
regions of recent star formation.  In the majority of the cases three
frames were obtained on each position in order to deal with cosmic
rays.

\placefigure{fig1}
\placefigure{fig2}

The data reduction was done using the IRAF\footnote{IRAF is
distributed by NOAO, which is operated by the Association of
Universities for Research in Astronomy Inc., under contract with the
National Science Foundation.} software package.  All the images were
bias subtracted, flat-fielded using normalized continuum lamps, and
then the frames for each slit position were combined to produce the
final images.  Because some of the H\,{\sc ii} regions we observed are
very faint, the extraction of their spectra was done with great care.
First, we extracted the spectrum of a standard star with a very strong
flux and used this image as a reference for the tracing of the spectra
of the fainter H\,{\sc ii} regions.  Finally, for the background
subtraction we used samples of sky as close as possible to the H\,{\sc
ii} regions, fitting the level of this background (night sky plus the
background light of NGC 1427A) with a low-order polynomial.

The wavelength calibration was done using He-Ne lamps taken just
before or just after each exposure.  We identified 23 good lines, with
which we constructed dispersion solutions with a fifth-order Legendre
polynomial, always obtaining a residual RMS of less than 0.1 $\AA$.
To measure the radial velocities we first fitted the continuum and
then subtracted it from every spectrum, so we are finally left with
just the emission lines.  By far the strongest line in all our spectra
is H$\alpha$.  This, along with the fact that there were far more
comparison lines, evenly distributed, on the red side than the blue
side of the wavelength range, making the dispersion solution better on
the red side, led us to use just the H$\alpha$ emission to measure the
velocities.  Once the continuum was subtracted, the velocities were
measured by fitting a Gaussian profile to the line to obtain the
center, and with this center we obtained the radial velocity by using
the standard Doppler formula.  Finally we applied the heliocentric
correction to all the H$\alpha$ velocities.

\placetable{tbl-1}  

To estimate the errors in our velocities we extracted various sky
spectra from each of the final images, selected ten to twelve night
emission lines at various signal-to-noise ratios (defined as S/N =
$f/(f+n(ron)^{2}+n(sky))^{1/2}$, where $f$ is the flux in electrons
contained in the emission line after the continuum was subtracted,
$sky$ is the continuum level at the emission line in electrons per
pixel, $n$ is the width of the line at zero intensity in pixels, and
$ron$ is the readout noise in electrons/ADU), and measured the centers
of all these lines following the same procedure as for the H$\alpha$
velocities.  Then we plotted the difference between each measurement
and the average of all the measurements of the same line (which we
call the residual) versus S/N.  In total we had approximately 1300
data points, which we grouped in bins and plotted.  The results are
shown in Fig. 3a.  To assign the error to a velocity, we measure the
S/N of the corresponding H$\alpha$ line and interpolate using the
diagonal rational function (\cite{pre92}) at that S/N.  The data are
given in Table 1, where the coordinates refer to the axes shown in
Fig. 1 and Fig. 5, and the origin is not included in the images.

\placefigure{fig3}

We performed another method of error estimation by means of a Monte
Carlo simulation.  We constructed an artificial spectrum consisting of
one perfectly gaussian emission line (i.e., we exactly know its
center, width and amplitude) placed in the wavelength region where we
observe the H$\alpha$ line in our spectra.  Then, using IRAF routines,
Poisson noise was added randomly to the `perfect' spectrum (using the
same gain, 0.8 electrons/ADU, and read-out noise, 3 electrons, as the
chip used during the observations), creating one thousand `noisy'
spectra.  Next, we added to these artificial spectra real sky randomly
extracted from the regions of NGC 1427A where no H\,{\sc ii} regions
are present.  Finally, we applied to these semi-artificial spectra the
same measuring process as for the H$\alpha$ velocities.  A histogram
of all the measurements fits well with a Gaussian curve (which tells
us that the measurement errors are normally distributed, an important
point when discussing modeling of the velocity field, see Section 3),
whose standard deviation we took as the error estimated for a
representative S/N of all the artificial spectra.  We automatized this
whole procedure and repeated it for many different S/N ratios,
obtaining results quite similar to those obtained with the analysis of
the skylines.  We chose then to adopt the night skylines method as our
error estimation.

\section{Kinematic models.}

In Fig. 4 we show the measured H$\alpha$ heliocentric velocities of
the 29 positions over the galaxy for which we measured reliable data.
The velocities are plotted as a function of the distance to the axis
of rotation, whose position was obtained as we will explain later in
this section.  On local scales, the data show a state of complex
kinematics, with very close points whose velocities do not overlap
within their error bars.  It is not uncommon for these clumpy Irr
galaxies to show disordered patterns in their velocity fields
(e.g. \cite{hun86}), but the large scatter of velocities observed in
NGC 1427A and the particular characteristics of its environment make
us suspicious about treating it as a normal Irr.  On a global scale,
one can see that there is a rotation present, with an amplitude of
about 150 km/s from one side of the galaxy to the other.  Most Irr
galaxies, unlike spirals (which usually show amplitudes in the
rotation speeds of 400 km/s from end to end), are slow rotators,
showing near rigid-body behaviour extending over most of their optical
dimensions (e.g. \cite{gal84}; \cite{hun86}).  In NGC 1427A it is
clear that the velocity rises from east to west following a roughly
linear trend (corresponding to solid-body rotation), $but$ $with$ $a$
$large$ $scatter$ $between$ $the$ $data$ $points$ $and$ $the$ $fitted$
$line$ (see Fig.4).  It is clear that any smooth, conventional model
of rotation curve will not be capable to follow such a large scatter.
However, trying to adjust some simple models to the data will uncover
overall characteristics and give some insights about the nature of the
velocity field of the galaxy.

\placefigure{fig4} 

As a first approximation, we tried to fit a rigid-body rotation model,

\centerline{\( v_{l.o.s.} = v_{0} + (\bf \omega \bf \times \bf r) \bf \cdot
(-\bf z) = \rm v_{0} + \omega_{y}x - \omega_{x}y, \)}

\noindent where $v_{0}$ is the recession velocity of the (arbitrary)
origin of the x-y coordinates on the plane of the sky (see Table 1),
$(-\bf z)$ is a unit vector along the line of sight, and $\omega_{x}$
and $\omega_{y}$ are the components of the angular velocity vector
$\bf \omega$ projected on the plane of the sky ($X$ being the E-W
direction and $Y$ the N-S one, see Fig. 4).  This model does not yield
any information concerning the center of rotation nor an inclination
of the disk of the galaxy.  A linear least-squares fit gives a
best-fit model with $\chi^{2}$=134, and a reduced $\chi^{2}$, or
$\chi^{2}$ per degree of freedom, of $\chi^{2}/(N-M)$=5.2, where
$N$=29 is the number of data points, and $M$=3 the number of
parameters to adjust.  This is a large value for the merit function
that is not acceptable in order to adopt the model as a good one.
However, the results are still valid as a first approximation to the
magnitude and direction of the rotation.  The best-fit model
parameters obtained were 1.29$\pm$0.05 and -12.8$\pm$0.1 km/s/kpc for
$\omega_{x}$ and $\omega_{y}$ respectively, and one can see that, as
expected from simple inspection, the rotation projected on the sky is
almost entirely around the N-S axis.  These values for the components
of the angular velocity vector imply an axis of rotation on the plane
of the sky whose direction is inclined $6^{\circ}$ counter-clockwise
from the vertical direction.  The shallow velocity gradient of about
13 km/s/kpc is in agreement with what is observed in most Irrs, having
$\approx$ 5-20 km/s/kpc (\cite{gal84}).
   
Next, we used a model after de Zeeuw and Lynden-Bell (1988), which
assumes that the gas lies in a flat disk following circular orbits.
The model represents a family of rotation curves, parametrized by
 
\[v_{rot}(r') = \frac{Vr'}{(r'^{2}+r_{0}^{2})^{p/2}}. \]

\noindent Here, $V$, $r_{0}$, and $p$ are constant parameters, and
$r'$ is the distance from each point to the center of rotation
measured on the plane of the galaxy. Note that the solid-body ($p$=0),
flat ($p$=1), Keplerian ($p$=3/2), and other models of rotation curves
are special cases of this family.  To allow for an arbitrary
inclination of the disk of the galaxy with respect to the sky, we did
the following.  First, using the center of rotation $(x_{0},y_{0})$ as
origin of coordinates, we rotated the $X$ and $Y$ axes (i.e., the
plane of the sky) by an angle $\beta$ around the line of sight $Z$,
obtaining the system $X''Y''Z''$, with $Z''= Z$.  After the fitting,
this angle will tell us about the direction of the axis of rotation
projected onto the sky.  Then we made a second rotation, now around
the $X''$ axis, tilting the $X''Y''$ plane by an angle $\alpha$,
obtaining the system $X'Y'Z'$.  The disk lies in the $X'Y'$ plane, and
the $Z'$ axis is parallel to the angular momentum vector of the
rotating disk.  The angle $\alpha$, then, sets the inclination of the
galaxy with respect to the plane of the sky.  In order to fit this
model to our data, we project the velocity of rotation along the line
of sight (the $-Z$ direction), so the equation to fit is

\[v_{l.o.s.} = v_{0} + v_{rot}(r') \sin\alpha\cos\theta'. \]

\noindent Here $v_{0}$ is the systemic velocity of the galaxy, and
$\theta'$ is the angle between the position vector $\bf r'$ and the
$X'$ axis.  This equation depends on eight parameters $(v_{0}, \beta,
\alpha, V, x_{0}, y_{0}, r_{0}, p)$, the majority of them in a
nonlinear way.  We developed a code that, using the
Levenberg-Marquardt method of nonlinear fitting (\cite{pre92}),
returns the values for the parameters that minimize the $\chi^{2}$
merit function.

\placefigure{fig5} 

In terms of the final value of $\chi^{2}$, the best-fit de Zeeuw $\&$
Lynden-Bell model resulted closer to the data than the pure rigid-body
rotation, but it is still not a good fit.  A careful inspection of
each step during the process of iteration to the best-fit model shows
that the parameters $v_{0}$, $\beta$, $\alpha$, $x_{0}$, and $y_{0}$,
quickly converge to their final, best-fit values.  The best-fit value
for the systemic velocity $v_{0}$ is 2039 km/s, in reasonable
agreement with the H\,{\sc i} systemic velocity measured by Bureau et
al. (1996).  The best-fit center of rotation, shown as a cross in
Fig. 5, is located approximately 12 arcsec to the west of the midpoint
between the optical edges of the galaxy.  We obtained an angle $\beta
= 10^{\circ}$, counter-clockwise from the N-S direction (see Fig. 5),
close to the inclination found using the solid-body model.  For the
inclination of the disk with respect to the sky, $\alpha$, the
best-fit value was $80^{\circ}$, which would correspond to a disk seen
almost edge-on (see Section 4 for the implications of this high
inclination).  However unexpected, this value of $\alpha$ is reached
quickly by the algorithm.  It does not agree with the inclination
reported by Bureau et al. (1996) of $\alpha = 48^{\circ}$, derived
using the photometric axial ratio, a rather arbitrary criterion for a
galaxy like NGC 1427A.  Having arrived at this point of the fitting
procedure, the merit function reaches a flat valley in parameter
space, with $\chi^{2} \approx$ 80, and $\chi^{2}$ per degree of
freedom of $\approx$ 3.8\footnote{For comparison, if we fix the value
of $\alpha = 48^{\circ}$ and run the fitting program, we obtain a
minimum $\chi^{2}$=122.}.  The parameters $p$, $V$, and $r_{0}$ are
degenerate, in the sense that there is no unique set that gives a
global minimum of $\chi^{2}$.  It is clear from the expression for
$v_{rot}$ that, as $p$ increases, $V$ also has to rise in order to
keep $v_{rot}$ constant.  This is indeed what the fitting algorithm
shows.  Setting $p=1$, we find $V=-75$ (in km/s only for this value of
$p$, and the sign indicating the direction of the spin), $r_{0}$=2.7
kpc, and $\chi^{2}$=83.  For $p=1.2$, $V=-220$, and $r_{0}=2.9$ kpc,
we have $\chi^{2}=81$.  And finally, for $p=1.5$, $V=-1175$, and
$r_{0}=3.7$ kpc, the $\chi^{2}$=79, almost negligibly better than the
model with $p=1$.  For values of $p<$1 the obtained $\chi^{2}$'s begin
to rise quickly.  As one can see, it is not possible to distinguish
between models with $1\leq p\leq 3/2$, because the data are scarce at
distances from the center where the models begin to differ from each
other.  This is shown in Fig. 6, where the data have been projected
onto the plane of the disk (dividing the corresponding velocities by
$\sin\alpha\cos\theta' $), and the corresponding error bars rescaled
(note that we didn't take the error in the scaling factor into
account).  Data points marked as triangles lie very close to the axis
of rotation, making the factor $\cos\theta '$ very small and uncertain
in its sign.  This is reflected by meaningless values of $v_{rot}$,
possibly with the wrong sign, as well as very large (but still
underestimated) error bars.

\placefigure{fig6}  

We explored the possibility that the models fail to explain the data
due to problems with our error estimation.  Of course, if we multiply
the errors in the measured velocities by a factor of, say, 1.5, then
the resulting value of $\chi^{2}$ would be acceptable.  But we are
quite confident that our quoted errors are not underestimated, having
obtained them by means of two different methods, one of them based on
the actual set of data.  Also, one could obtain a bad fit, even with
the correct model, if the quoted errors were not normally (Gaussian)
distributed.  The reason for this is that the minimization of the
merit function $\chi^{2}$ $assumes$ that the errors are normally
distributed (\cite{pre92}).  We tested this possibility by building
(for both methods of error estimation) the distribution of the errors
and fitting a Gaussian function to them.  In both cases the agreement
was very good, as seen in Fig. 3b for the method using skylines, so we
reject this possibility.  All the previous attempts to fit a model to
the data and the above discussion about the distribution of the errors
was regarding the estimated errors in the velocities.  So far we have
assumed that we know $exactly$ the positions $(x,y)$ of the regions
whose spectra we have.  So, in order to quantify the changes in the
fitting results when some uncertainty in the coordinates is
introduced, we performed the following exercise.  We took the
coordinates $(x,y)$ of all the data points and changed randomly their
values around the original ones, after which we adjusted the
rigid-body model to the ``new'' data set.  We estimate a ``real''
uncertainty in the coordinates to be no more than 2 arcsec in the
worst of the cases, so the changes introduced in the coordinates were
randomly distributed between plus or minus 2 arcsec.  Repeating the
procedure two or three thousand times, always in a random way, we
found that $\omega_{x}$, $\omega_{y}$, and $\chi^{2}$ never change by
a large amount.  Doubling the uncertainty to 4 arcsec does not make
any difference, so we conclude that there is no need to worry about
uncertainties in the coordinates.  Finally, since we are interested in
relative velocities, eventual systematic errors should not affect our
results as long as they affect all velocities in the same way.

\section{Discussion}
 
\subsection{Kinematics} 

We have presented the velocities of the ionized gas from many of the
brightest H\,{\sc ii} regions in NGC 1427A, and modeled them to derive
the basic properties of its dynamics.  Using two different models for
the kinematics we found the major axis of rotation, with both
solutions in reasonable agreement.

The simplest model, a global rigid-body rotation plus a random
component on small scales (responsible for the poor fit), seems to be
a good approximation to the data (Fig. 4), and is in concordance with
what is observed in most Irr's.  The radial velocities of points in
the North Object match well with this model (see Fig. 4), which
suggests that it is part of the galaxy, as the rest of the H\,{\sc ii}
regions.  However, if we want information about the center of rotation
and the inclination of the galaxy, we need a more elaborate model.
Our solution using the de Zeeuw $\&$ Lynden-Bell model is better than
the solid-body one in terms of the merit function $\chi^{2}$ but, here
again, the random component dominates the appearance of the rotation
curve (Fig. 6).  The puzzling feature of this solution is the
remarkably high inclination ($80^{\circ}$) returned by the fit, which
does not depend on whether we use the points in the North Object in
the fitting procedure.  Assuming that the North Object is part of NGC
1427A and that it lies in the same disk as the rest of the H\,{\sc ii}
regions, this inclination would place it at a distance of about 30 kpc
from the fitted (and optical) center of NGC 1427A, which is difficult
to believe.  If we lower the angle of inclination until the one
derived using the photometric axial ratio (\cite{bur96}) this problem
is softened, with the North Object at 8.2 kpc, but with a $\chi^{2}$
50\% higher than before.  So, we are inclined to place the North
Object outside the disk of NGC 1427A.  We estimated the probability of
the chance coincidence that the North Object being an independent
cluster member with its velocity in the same range as those of the
H\,{\sc ii} regions of NGC 1427A (1950-2100 km/s).  Assuming for the
cluster galaxies a Gaussian radial velocity distribution (which is the
case when the three dimensional distribution is Maxwellian) centered
at NGC 1399 (1430 km/s) and with a dispersion of 325 km/s
(\cite{bur96}), we obtain a probability of 3.5\% of a chance
coincidence.  This low probability, the North Object lying outside the
plane of the galaxy, and the coincidence in the radial velocities
would indicate that it is a separate object but gravitationally bound
to NGC 1427A, probably a small satellite orbiting the galaxy.

Based on the previous results, we estimated the dynamical mass and
other related quantities for NGC 1427A.  Taking the angular velocity
obtained from the solid-body fit and assuming a spherical mass
distribution, the total mass inside a radius of 6.2 kpc (the size of
the major axis at the 24.7 mag/arcsec$^{2}$ isophote in V) is $M_{dyn}
\gtrsim$ (9 $\pm$ 3)x$10^{9} M_{\sun}$
\footnote{The uncertainty in the total mass is almost entirely due to
the uncertainty in the size of NGC 1427A, which is a combination of
uncertainties in the angular size of the galaxy and the distance to
Fornax.}.  This is a lower limit for the total mass inside this radius
because of the unknown component of the angular velocity along the
line of sight.  However, if the inclination of the disk is really as
high as $80^{\circ}$, then this unknown component will not be very
relevant, and the quoted value for $M_{dyn}$ will be close to the
actual one.  The mass in the form of neutral hydrogen can be obtained
from the integrated H\,{\sc i} flux (\cite{bur96}) and the adopted
distance to Fornax, using the formula of Roberts (1975).  With this,
the H\,{\sc i} mass turns out to be $M_{H\,{\sc i}}$ = (1.8 $\pm$
0.3)x$10^{9} M_{\sun}$, so the fraction of the total mass in the form
of neutral hydrogen is approximately 0.2, twice the value for the LMC
(based on the total mass from \cite{kun97} and the H\,{\sc i} flux
from \cite{huc88}).  Finally, from the magnitudes given by Hilker et
al. (1997), the mass-to-light ratios for NGC 1427A are $M/L_{B}
\gtrsim$ 3.9, and $M/L_{V} \gtrsim$ 4.8, in units of solar masses per
solar luminosities in the corresponding band.  As a comparison, the
LMC has a mass-to-light ratio of $\approx 2.9 M_{\sun}/L_{\sun}$ in
the B band (from the magnitudes given by \cite{dev91} and the total
mass of \cite{kun97}).  The values obtained for the total mass of NGC
1427A, the fraction of H\,{\sc i} in it, as well as the mass-to-light
ratio, all are in good agreement with typical values for the latest
galaxy types, as summarized by Roberts \& Haynes (1994).

The random behaviour on small scales is not difficult to understand.
Since the aperture sizes of our spectra vary between 4 and 12 arcsec,
corresponding to spatial extensions in the range 0.3 - 1 kpc on the
galaxy, our velocities are actually averages taken over structures and
regions of various sizes.  On these scales it is very common to find
in these galaxies structures such as shells and supershells,
large-scale filaments of ionized gas, as well as a non-negligible
component of diffuse ionized gas (\cite{hun86}; \cite{mar97}, 1998).
All these structures reflect the strong impact that massive stars have
on their surroundings, injecting large amounts of energy via stellar
winds and supernova shocks.  Hints of two supershells can be seen in
the optical images of NGC 1427A, with diameters of 0.7 and 1.1 kpc
(see Fig. 7), apparently emerging from the largest of the high surface
brightness features.  These structures seem to be primarily
photoionized, despite their location very far from the nearest star
associations (\cite{hun97}; \cite{mar98}), and show expansion
velocities between 20 and 60 km/s, sometimes going up to 100 km/s (see
Fig. 3 in \cite{mar98}).  The filled circles in the rotation curve of
Fig. 4 correspond to the brightest H\,{\sc ii} regions seen in Fig. 5,
and one can see that they are closer to the solid-body line than the
blank circles, which correspond to diffuse ionized gas some distance
away from the bright H\,{\sc ii} regions.  This diffuse gas should be
more subject to the effects of expanding shells and filaments, and
this could be the reason why they depart from the overall rotation.
The largest discrepancies in our data are between 40 and 70 km/s, so
it is very likely that some of them are due to the strong influence of
very massive stars on the ISM.  Furthermore, part of the diffuse gas
may not be in the disk of the galaxy, but instead it could have been
transported into the halo by some mechanism (see, e.g., Dahlem,
Dettmar, \& Hummel (1994) for ionized gas away from the disk in NGC
891, and also Bomans, Chu, \& Hopp (1997) for gas outflows from
intense star forming regions in NGC 4449), where it would not
necessarily corotate with the disk.  Nevertheless, considering only
the bright H\,{\sc ii} regions does not improve the fits (the
dispersion is smaller, but so are the error bars).  Therefore, some
physical mechanism (winds, turbulence, ...)  must still be involved to
explain the $\approx$ 10-15 km/s discrepancies.

\placefigure{fig7} 

\subsection{Interaction with the cluster environment}

Hilker et al (1997) and Cellone $\&$ Forte (1997) already suggested,
based on morphological reasons and colours of the H\,{\sc ii} regions,
that the appearance of NGC 1427A is due to an interaction with the
Fornax Cluster environment.  This possibility is very likely given the
location of NGC 1427A near the center of the cluster.  Based on the
obvious alignment of the bright giant H\,{\sc ii} regions along a half
ring at the south western part of the galaxy and the colors of the
only two bright knots at the extreme north (``the North Object''),
Cellone $\&$ Forte suggested that this could be the encounter between
two different objects, the North Object being one of the many dwarf
ellipticals that populate the center of the Fornax Cluster.  As we
said before, assuming a solid-body rotation, the velocities for the
North Object fall well into the general kinematical pattern, which
would indicate, with high probability, that it is just another part of
NGC 1427A, not an intruder galaxy.  On the other hand, if we take the
de Zeeuw \& Lynden-Bell model as the valid one, then we would have to
accept that the North Object is not in the same disk as the rest of
the H\,{\sc ii} regions, possibly being a satellite galaxy of NGC
1427A.  The proximity of the two giant ellipticals of the cluster, NGC
1399 and NGC 1404, suggests that NGC 1427A might be experiencing
strong tidal forces.  Tidal interaction is also a proposed mechanism
for triggering star formation, but it seems unlikely that this could
produce the ring-like pattern of star forming regions along one edge
of the galaxy.  Tides are known to produce thin low surface brightness
filaments that stretch out from interacting galaxies (\cite{gre98}).
A search for tails at this low surface brightness would be possible
with the use of wide-field imaging plus relatively large pixel sizes
(in order to collect more light at the expense of resolution).

We argue here that the most likely scenario to explain the
morphological and kinematical features of NGC 1427A is its passage
through the hot ICM of Fornax.  When a galaxy crosses the ICM of a
cluster at a supersonic speed, a shock front will appear before the
galaxy.  This will abruptly raise the temperature and density of the
ICM gas that goes through it, and so, behind the shock, the galaxy
will be exposed to the action of a high thermal pressure plus the ram
pressure that the shocked intracluster gas exerts upon it.  Given the
small sound speeds in the interstellar gas, it is very likely that
another shock will form, now inside the galaxy.  If the shocked
interstellar gas has a cooling time\footnote{The cooling time is
estimated by $t_{cool} \approx (3/2)kT/n\Lambda$, where $\Lambda$ is
the volume emissivity of the gas divided by the electron density and
proton density (the cooling coefficient).  We adopt the cooling curve
of Gehrels and Williams (1993).}  much shorter than the time needed by
the shock wave to cross the medium, it will cool very rapidly, with
the subsequent condensation that pressure equilibrium requires.  In
this way, dense shells of cold material follow immediately behind this
$`isothermal$ $shock$' (also called a radiative shock).  Molecular
clouds are formed when the column density of these cold clouds exceeds
the threshold at which UV dissociation is truncated (\cite{fra86}),
and when parts of these dense shells are fragmented and become
gravitationally unstable (see \cite{elm78}) new stars are formed.
This is how regions of active star formation may align around the
edges of gas-rich cluster galaxies, as in the galaxies observed by
Gavazzi et al. (1995).

NGC 1427A is at a projected distance of 120 kpc from NGC 1399 and
moving at a relative radial velocity $V_{r} \approx$ 600 km/s
(\cite{bur96}), so it will be in contact with the densest parts of the
ICM during $t_{ICM} \approx$ 2x$10^{8}$ years, a time long enough to
allow shocks propagate into the ISM and trigger new star formation.
Note that, since NGC 1427A is a gas-rich galaxy, it is probably
crossing the Fornax ICM for the first time.  The X-ray emitting plasma
in Fornax has a temperature of 1.3x$10^{7}$ K (\cite{ran95}), and a
density of $\approx 10^{-3}$ cm$^{-3}$ at the distance of NGC 1427A
(\cite{ike96}).  The adiabatic sound speed in a completely ionized
medium with temperature $T$ is $c_{s}$ $\approx$ 0.15 $T^{1/2}$ km/s
\footnote{We assume a gas with primordial abundances (90\% hydrogen
and 10\% helium in number) so, for complete ionization the mean
molecular weight is 0.59, and with just singly ionized helium it would
be 0.61.}  , which for the ICM in Fornax gives $c_{ICM} \approx$ 500
km/s.  If we assume that this hot intracluster gas moves with NGC 1399
(around which it appears to be centered, see Fig.1 in \cite{jon97}),
then the passage of NGC 1427A across the ICM is supersonic, with an
approximate Mach number $M \approx$ 1.2 (a lower bound, since we only
know one component of the relative velocity).  A weak adiabatic shock
will be leading the way of NGC 1427A through the ICM, slightly raising
the temperature and density of the gas that crosses it.

The ISM in gas-rich galaxies is extremely complex, with the
thermodynamic properties of the different phases varying rapidly from
place to place and also in time (see, e.g., \cite{kul88} for a
discussion of the Milky Way's ISM; and also \cite{mck77}).  We will
discuss the situation for two representative states of the ISM: a
hypothetical hot ionized halo, and a warm neutral hydrogen disk.  In
order to keep the halo in hydrostatic equilibrium in the galaxy's
potential well as revealed by its rotation curve, the required
temperature of this hypothetical gas is $\approx$ 2x10$^{5}$ K.  At
this temperature the sound speed is $c_{halo} \approx$ 70 km/s.
Taking the observed mean value for the pressure of the Milky Way's ISM
of $<P_{ISM}>$ $\approx 3000$ cm$^{-3}$K (\cite{kul88}), we would have
a halo density of 1.5x$10^{-2}$ cm$^{-3}$.  Note that, for these
conditions, the cooling time is $\approx$ 6x$10^{5}$ years, so
constant energy input is required to keep the gas at this temperature.
Assuming that most of the incident momentum from the ICM is
transferred to the galaxy, we obtain $v_{ISM} \equiv v_{halo} \approx
(\rho_{ICM} /\rho_{halo})^{1/2} v_{ICM} \approx$ 150 km/s.  Then,
there would be a shock with $M \approx$ 2.  Applying the
Rankine-Hugoniot jump conditions (\cite{lan79}) we obtain behind this
shock a temperature of $\approx$ 4x$10^{5}$ K and a density of
$\approx$ 3.5x$10^{-2}$ cm$^{-3}$.  With these values, the cooling
time for the shocked gas in the halo would be slightly $larger$ than
the cooling time before the shock appeared.

For our H\,{\sc i} phase, we may take an original temperature of
$10^{4}$ K
\footnote{This temperature is at the higher end of the observed range
for this gas phase in the Galaxy, but we adopt it because at lower
temperatures the cooling function is uncertain due to the varying
degree of ionization.  However, the conclusions will be the same as
long as, below 10000 K, the slope of the cooling curve remains
positive.}.  Then the sound speed in this medium will be $c_{H\,{\sc
i}} \approx$ 10 km/s (here the mean molecular weight is 1.23 if
everything is neutral), and using $<P_{ISM}>$ the density would be 0.3
cm$^{-3}$.  Again, the cooling time is short, so constant energy input
is required.  With these values, the velocity of the shock within the
H\,{\sc i} medium turns out to be $v_{ISM} \equiv v_{H\,{\sc i}}
\approx$ 30 km/s, and now we have a shock with $M \approx$ 3.  Using
the jump conditions we have that behind the shock the temperature of
the H\,{\sc i} is 4x$10^{4}$ K and its density 1 cm$^{-3}$.  The
cooling time of the shocked H\,{\sc i} would be $t_{cool-H\,{\sc i}}
\approx$ 3000 years, more than $20$ $times$ $shorter$ than the cooling
time for the unperturbed H\,{\sc i}.  The reason for this is that in
the range of temperatures for the H\,{\sc i} phase the cooling
function has a positive slope, while in the range of temperatures of
the halo gas this slope in negative (see Fig.1 in \cite{geh93}).

The halo shock takes $\approx$ 4x10$^{7}$ years to fully cross a
spherical halo with a radius of 6 kpc, while the shock in the neutral
phase would need 3x10$^{7}$ years to move just 1 kpc.  In both media,
the cooling time is much shorter than the shock crossing time, so we
could regard them as isothermal shocks.  However, in the halo (where
the cooling time scales before and after the shock are of the same
order), this consideration does not apply if the agents that
originally kept the gas at its equilibrium temperature are still
present regardless of the shock, so the gas would be unable to cool.
If this is not the case, all the halo gas accumulated behind the shock
will cool and eventually be detected as H\,{\sc i}.  In the H\,{\sc i}
disk, the swept up gas behind the shock will surely cool rapidly, form
molecular clouds, and trigger bursts of new star formation.

The rotation rate of $\approx$ 13 km/s/kpc (a lower bound, since we do
not know the component of the angular velocity along the line of
sight) corresponds to a rotation period of $T \approx$ 4.5x$10^{8}$
years, comparable to the crossing time, $t_{ICM}$, and much longer
than the lifetimes of normal H\,{\sc ii} regions, $t_{H\,{\sc ii}}
\leq$ 10-15 Myr (given by the lifetimes of the very massive stars
whose ionizing fluxes generated them in the first place).  Thus, it is
not surprising that these star forming complexes are only found along
one side of the galaxy, which would have to be the side directly
exposed to the shocked ICM.  This explains the bow-shock appearance of
the south-western edge of NGC 1427A, since the H\,{\sc ii} regions
formed at the interacting side do not last long enough to reach the
other side, following the rotation of the galaxy.  The same scenario
was proposed by de Boer et al (1998) for the interaction between the
LMC and the hot Milky Way halo, giving as evidence for it the
existence of a gradient in the ages of the peripheral young star
clusters of the LMC in the direction expected from the relative motion
between both galaxies.  To obtain this kind of evidence is obviously
not possible in the case of NGC 1427A because we can not resolve the
young star clusters behind the H\,{\sc ii} regions at this distance.

\section{Conclusions} 

We have obtained the ionized gas kinematics of NGC 1427A by means of
long slit spectroscopy of the brightest H\,{\sc ii} regions.  The
velocity field follows, on average, solid body rotation over the whole
optical dimensions.  Looking closer, however, there are large
discrepancies in some data points, most of them associated with the
diffuse component of the ionized gas in regions far away from the
center of rotation.

We modeled the kinematics using two models of rotation, both assuming
circular orbits in a flat disk.  There is agreement between both
models regarding the inclination of the axis of rotation, which is
near the N-S direction.  The rigid-body fit gives an angular velocity
of 13 km/s/kpc, which is consistent with what is observed in this type
of galaxies.  The de Zeeuw and Lynden-Bell model fits the data better
than the simpler solid-body but yields an unexpectedly high
inclination ($\approx 80^{\circ}$) of the disk of the galaxy.  Both
models give large values for the merit function $\chi^{2}$ because the
set of velocities shows a random component that is important on small
scales.  This behaviour alone does not provide evidence for an
interaction with the cluster environment, and may be explained by the
impact that massive stars has on the ISM in Irr galaxies.

We reject the scenario in which NGC 1427A is the result of a collision
with a smaller member of the cluster, because the only candidate
intruder, the North Object, has a radial velocity which is nicely
coincident with the general velocity pattern.  However, if the
inclination of the disk derived from the de Zeeuw and Lynden-Bell
model is adopted, we can not place the North Object in the same disk
as the rest of the H\,{\sc ii} regions.  Instead, it would turn out to
be a small satellite of NGC 1427A.

Several properties of NGC 1427A and its environment strongly suggest
that this galaxy is interacting with the hot gas that pervades the
cluster center, and we are inclined to favor this scenario.  We have
given quantitative estimates (although some of the numbers we used are
just reasonable guesses) in order to show how the bow-shock alignment
of the recent star formation in NGC 1427A is very likely due to the
ram pressure from the ICM of Fornax as the galaxy crosses it.  Further
evidence for this scenario will have to wait for more detailed
kinematics, such as interferometric Fabry-Perot imaging and good
resolution stellar spectra.  Then it will be possible to compare the
kinematics of the gas component with that of the stars, which may be
very different in the ram pressure scenario.  Also, high resolution
mapping in H\,{\sc i} should show signs of this interaction, such as
stripped gas and sudden truncation and asymmetries in the distribution
of the neutral gas, as observed in the Virgo Cluster (\cite{cay90})
and even in groups of galaxies (\cite{dav97}).

$Acknowledgements$ We thank Bill Kunkel for allowing us to use part of
his observing time, and without him this work wouldn't have started.
We also thank Guillermo Tenorio-Tagle for invaluable discussions and
insights; Mar\'\i a Teresa Ruiz and Michael Hilker for their interest
and help in the continuation of this work; and Roberto Terlevich for
useful comments.  We thank to Fondecyt Chile for support through
``Proyecto FONDECyT 8970009''.



\begin{deluxetable}{crrrrrrrrrrr}
\footnotesize
\tablecaption{The data. \label{tbl-1}}
\tablewidth{0pt}
\tablehead{
\colhead{label} & \colhead{$x$} & \colhead{$y$} & \colhead{$v_{helio}$} & \colhead{$\Delta v$} \\ & \colhead{(pix)} & \colhead{(pix)} & \colhead{(km/s)} & \colhead{(km/s)}
}
\startdata
1  &343.2  &441.13  &1955.1  &10.2  \nl
2  &238.8  &441.13  &2034.2  &5.26  \nl
3  &209.8  &441.13  &2042.6  &3.62  \nl
4  &161.8  &441.13  &2038.6  &5.66  \nl
5  &143.6  &441.13  &2047.9  &9.32  \nl
6  &348.6  &432.56  &1983.9  &5.07  \nl
7  &237.2  &432.56  &2019.7  &5.47  \nl
8  &215.6  &432.56  &2041.4  &5.06  \nl
9  &146.4  &432.56  &2054.1  &3.65  \nl
10 &231.9  &423.96  &2018.1  &4.21  \nl
11 &167.4  &423.96  &2046.8  &4.36  \nl
12 &147.0  &423.96  &2076.3  &3.74  \nl
13 &343.3  &415.36  &1984.2  &4.98  \nl
14 &236.2  &415.36  &2019.9  &5.09  \nl
15 &189.2  &415.36  &2030.9  &8.40  \nl
16 &151.3  &415.36  &2058.1  &5.12  \nl
17 &119.4  &415.36  &2086.9  &5.12  \nl
18 &88.9   &415.36  &2060.9  &9.45  \nl
19 &231.2  &421.73  &2016.2  &3.79  \nl
20 &205.7  &396.20  &2035.9  &4.49  \nl
21 &185.2  &377.20  &2056.3  &9.49  \nl
22 &173.7  &366.20  &2053.4  &9.59  \nl
23 &164.7  &353.70  &2032.3  &9.69  \nl
24 &125.4  &319.89  &2035.4  &11.3  \nl
25 &257.2  &249.90  &2033.4  &4.00  &(N.O.) \nl
26 &247.7  &243.70  &2024.3  &9.00  &(N.O.) \nl
27 &163.6  &324.70  &2045.8  &3.63  \nl
28 &127.0  &362.10  &2068.0  &9.26  \nl
29 &111.3  &379.50  &2108.4  &7.72  \nl

\enddata


\tablenotetext{a}{N.O. = North Object}

\end{deluxetable}

\clearpage

\figcaption[fig1.ps]{Isophotal contours of a B-band image of NGC 1427A
(\cite{hil97}) with the slit positions superimposed.  The North Object
(N.O.) is indicated with a circle.  The isophotal levels were chosen
in order to enhance the contrast between the H\,{\sc ii} regions and
the main stellar body.  The object at approximate coordinates (60,280)
is a background galaxy. \label{fig1}}

\figcaption[fig2.ps]{Sample spectra of H\,{\sc ii} regions in NGC
1427A at different signal-to-noise ratios (S/N) for the H$\alpha$
line. Units of the vertical axis are CCD counts. \label{fig2}}

\figcaption[fig3a.ps,fig3b.ps]{Error analysis.  (a) The uncertainty in
the center of the H$\alpha$ line as a function of the signal-to-noise
ratio.  An error of 0.1 $\AA$ translates to 4.5 km/s in velocity.  (b)
The distribution of residuals in the method of error estimation using
night skylines.  The solid curve is a Gaussian fit to the points.  It
is centered at 0.04 $\AA$ and has a standard deviation of 0.18 $\AA$.
\label{fig3}}

\figcaption[fig4.ps]{The radial velocities of the H$\,{\sc ii}$
regions in NGC 1427A vs. the distance to the axis of rotation found
with the solid-body model (6$^{\circ}$ counter-clockwise from the N-S
direction).  Note the agreement between the velocities of points in
the North Object (N.O.) with the overall rotation of the
galaxy. Filled circles represent the peaks of the H$\alpha$ emission,
and blank circles the diffuse emission at some distance from the
bright H$\,{\sc ii}$ regions. \label{fig4}}

\figcaption[fig5.ps]{V image of NGC 1427A (\cite{hil97}), showing the
alignment of H$\,{\sc ii}$ regions along half of its perimeter.  The
positions of the axis of rotation and the center of rotation found
with the de Zeeuw \& Lynden-Bell model are shown, as well as the
direction to the giant ellipticals at the cluster
center. \label{fig5}}

\figcaption[fig6.ps]{Circular velocities on the plane of the disk of
NGC 1427A, according to the best-fit model of de Zeeuw \& Lynden-Bell.
The triangles represent points very close to the axis of rotation.
The two points located at the North Object also are triangles, but
they are not shown because they would fall at $\approx$ 30 kpc.
Filled symbols are the peaks of the H$\alpha$ emission, and blank
symbols the diffuse emission.  Also shown are the rotation curves with
the values of $p$ between which the data can not
distinguish. \label{fig6}}

\figcaption[fig7.ps]{Detail of the V image of NGC 1427A, showing
possible large scale expanding shells in the interstellar medium.  The
arrows mark structures with diameters of 0.7 and 1.1
kpc. \label{fig7}}


\begin{thebibliography}{}

\bibitem[Barnes \& Hernquist 1991]{bar91} Barnes, J., \& Hernquist, L. 1991, \apj, 370, L65  

\bibitem[Bomans et al.\ 1997]{bom97} Bomans, D.J., Chu, Y.-H., \& Hopp, U. 1997, \aj, 113, 1678 

\bibitem[Bureau et al.\ 1996]{bur96} Bureau, M., Mould, J. R., \& Staveley-Smith, L. 1996, \apj, 463, 60

\bibitem[Butcher \& Oemler 1978]{but78} Butcher, H., \& Oemler, A., Jr. 1978, \apj, 219, 18 

\bibitem[Byrd \& Valtonen 1990]{byr90} Byrd, G., \& Valtonen, M. 1990, \apj, 350, 89

\bibitem[Cayatte et al.\ 1990]{cay90} Cayatte, V., van Gorkom, J. H., Balkowski, \& C., Kotanyi, C. 1990, \aj, 100, 604 

\bibitem[Cellone \& Forte 1997]{cel97} Cellone, S. A., \& Forte, J. C. 1997, \aj, 113, 1239 

\bibitem[Dahlem et al.\ 1994]{dah94} Dahlem, M., Dettmar, R.-J., \& Hummel, E. 1994, \aap, 290, 384

\bibitem[Davis et al.\ 1997]{dav97} Davis, D., Keel, W., Mulchaey, J., \& Henning, P. 1997, \aj, 114, 613

\bibitem[de Boer et al.\ 1998]{deb98} de Boer, K. S., Braun, J. M., Vallenari, A., \& Mebold, U. 1998, \aap, 329, L49

\bibitem[Della Valle et al.\ 1998]{del98} Della Valle, M., Kissler-Patig, M., Danziger, J., \& Storm, J. 1998, \mnras, 299, 267 

\bibitem[de Zeeuw \& Lynden-Bell 1988]{dez88} de Zeeuw, P. T., \& Lynden-Bell, D. 1988, \mnras, 232, 419

\bibitem[de Vaucouleurs et al.\ 1991]{dev91} de Vaucouleurs, G., de Vaucouleurs, A., Corwin, H. G., Buta, R. J., Paturel, G., \& Fouqu\'e, P. 1991, Third Reference Catalogue of Bright Galaxies

\bibitem[Elmegreen \& Elmegreen 1978]{elm78} Elmegreen, B. G., \& Elmegreen, D. M. 1978, \apj, 220, 1051 

\bibitem[Evrard 1991]{evr91} Evrard, A.E. 1991, \mnras, 248, Short Communications 8p-10p

\bibitem[Ferguson \& Sandage 1988]{fer88} Ferguson, H. C., \& Sandage, A. 1988, \aj, 96, 1520

\bibitem[Franco \& Cox 1986]{fra86} Franco, J., \& Cox, D. P. 1986, \pasp, 98, 1076 

\bibitem[Gallagher \& Hunter 1984]{gal84} Gallagher, J., \& Hunter, D. 1984, Ann. Rev. Astron. Astroph. 22, 37

\bibitem[Gavazzi et al.\ 1995]{gav95} Gavazzi, G., Contursi, A., Carrasco, L., Boselli, A., Kennicutt, R., Scodeggio, M., \& Jaffe, W. 1995, \aap, 304, 325  

\bibitem[Gehrels \& Williams 1993]{geh93} Gehrels, N., \& Williams, E. D. 1993, \apj, 418, L25 

\bibitem[Giovanell \& Haynes 1985]{gio85} Giovanelli, R., \& Haynes, M. 1985, \apj, 292, 404

\bibitem[Gratton et al.\ 1997]{gra97} Gratton, R. G., Fusi Pecci, F., Carretta, E., Clementini, G., Corsi, C. E., \& Lattanzi, M. G. 1997, in Hipparcos Venice'97 Symposium (ESA SP-402) 

\bibitem[Gregg \& West 1998]{gre98} Gregg, M. D., \& West, M. J. 1998, Nature, 396, 549  

\bibitem[Grillmair et al.\ 1999]{gri99} Grillmair, C., Forbes, D., Brodie, J., \& Elson, R. 1999, \aj, 117, 167

\bibitem[Gunn \& Gott 1972]{gun72} Gunn, J., \& Gott, J. R. 1972, \apj, 176, 1

\bibitem[Gunn 1989]{gun89} Gunn, J. 1989, in The Epoch of Galaxy Formation, eds. Frenk, C. S. et al. (Kluwer, Dordretch), 167  

\bibitem[Henriksen \& Byrd 1996]{hen96} Henriksen, M. J., \& Byrd, G. 1996, \apj, 459, 82

\bibitem[Hilker et al.\ 1997]{hil97} Hilker, M., Bomans, D. J., Infante, L., \& Kissler-Patig, M. 1997, \aap, 327, 562

\bibitem[Hilker 1998]{hil98} Hilker, M. 1998, PhD. Thesis, University of Bonn

\bibitem[Huchtmeier \& Richter 1988]{huc88} Huchtmeier, W. K., \& Richter, O. -G. 1988, \aap, 203, 237 

\bibitem[Hunter \& Gallagher 1986]{hun86} Hunter, D., \& Gallagher, J. 1986, \pasp, 98, 5

\bibitem[Hunter \& Gallagher 1997]{hun97} Hunter, D., \& Gallagher, J. 1997, \apj, 475, 65

\bibitem[Ikebe et al.\ 1996]{ike96} Ikebe, Y., Ezawa, H., Fukuzawa, Y., Hirayama, M., Ishisaki, Y., Kikuchi, K., Kubo, H., Makishima, K., Matsushita, K., Ohashi, T., Takabashi, T., \& Tamura, T. 1996, Nature, 379, 427 

\bibitem[Jensen et al.\ 1998]{jen98} Jensen, J., Tonry, J., \& Luppino, G. 1998, \apj, 505, 111

\bibitem[Jones et al.\ 1997]{jon97} Jones, C., Stern, C., Forman, W., Breen, J., David, L., Tucker, W., \& Franx, M. 1997, \apj, 482, 143 

\bibitem[Kohle et al.\ 1996]{koh96} Kohle, S., Kissler-Patig, M., Hilker, M., Richtler, T., Infante, L., \& Quintana, H. 1996, \aap, 309, L39  

\bibitem[Kulkarni \& Heiles 1988]{kul88} Kulkarni, S. R., \& Heiles, C. 1988, in Galactic and Extragalactic Radio Astronomy, ed. G. L. Verschuur \& K. I. Kellermann (Springer-Verlag), 95     

\bibitem[Kunkel et al.\ 1997]{kun97} Kunkel, W. E., Demers, S., Irwin, M. J., \& Albert, L. 1997, \apj, 488, L129 

\bibitem[Landau \& Lifshitz 1979]{lan79} Landau, L. D., \& Lifshitz, E. M. 1979, in Fluid Mechanics (Pergamon Press), 331 

\bibitem[Lee et al.\ 1997]{lee97} Lee, M. G., Kim, E., \& Geisler, D. 1997, \aj, 114, 1824


\bibitem[Lynds \& Toomre 1976]{lyn76} Lynds, R., \& Toomre, A. 1976, \apj, 209, 382

\bibitem[Martin 1997]{mar97} Martin, C. 1997, \apj, 491, 561

\bibitem[Martin 1998]{mar98} Martin, C. 1998, \apj, 506, 222

\bibitem[McKee \& Ostriker 1977]{mck77} McKee, C. F., \& Ostriker, J. P. 1977, \apj, 218, 148 

\bibitem[Moore et al.\ 1996]{moo96} Moore, B., Katz, N., Lake, G., Dressler, A., \& Oemler, A. 1996, Nature, 379, 613

\bibitem[Patterson \& Thuan 1992]{pat92} Patterson, R., \& Thuan, T. X. 1992, \apj, 400, L58 

\bibitem[Phookun \& Mundy 1995]{pho95} Phookun, B., \& Mundy, L. G. 1995, \apj, 453, 154

\bibitem[Press et al.\ 1992]{pre92} Press, W. H., Teukolsky, S. A., Vetterling, W. T., \& Flannery, B. P. 1992, in Numerical Recipes (Cambridge University Press)   

\bibitem[Rangarajan et al.\ 1995]{ran95} Rangarajan, F. V. N., Fabian, A. C., Forman, W. R., \& Jones, C. 1995, \mnras, 272, 665 

\bibitem[Richtler et al.\ 1992]{ric92} Richtler, T., Grebel, E. K., Domg\"{o}rgen, H., Hilker, M., \& Kissler-Patig, M. 1992, \aap, 264, 25 

\bibitem[Roberts 1975]{rob75} Roberts, M. 1975, in Galaxies and the Universe, ed. A. Sandage, M. Sandage, \& J. Kristian (Chicago Press), 326 

\bibitem[Roberts and Haynes 1994]{rob94} Roberts, M. S., \& Haynes, M. P. 1994, Ann. Rev. Astron. Astroph. 32, 115

\bibitem[Theys \& Spiegel 1977]{the77} Theys, J. C., \& Spiegel, E. A. 1977, \apj, 212, 616 

\end{thebibliography}
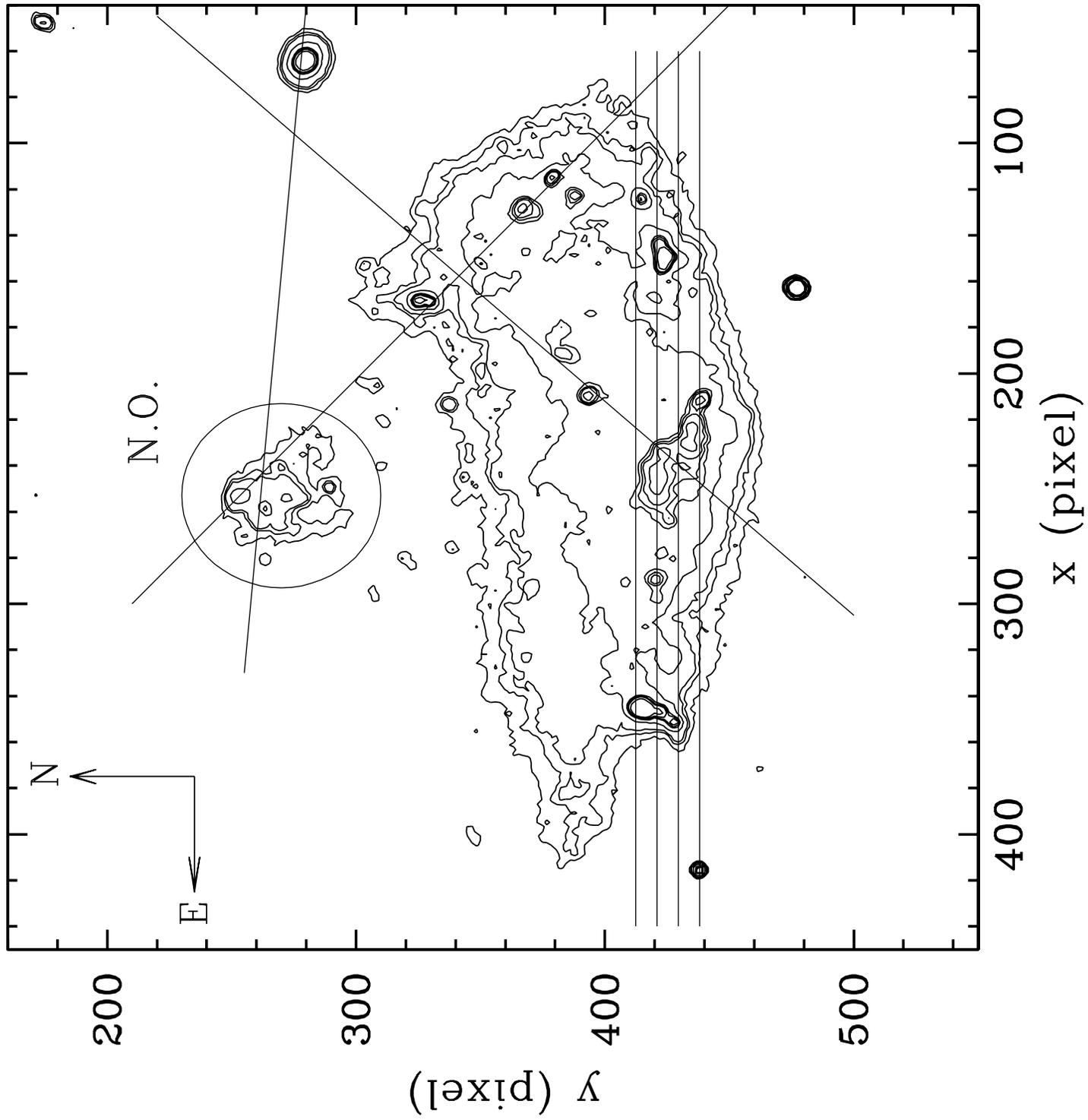
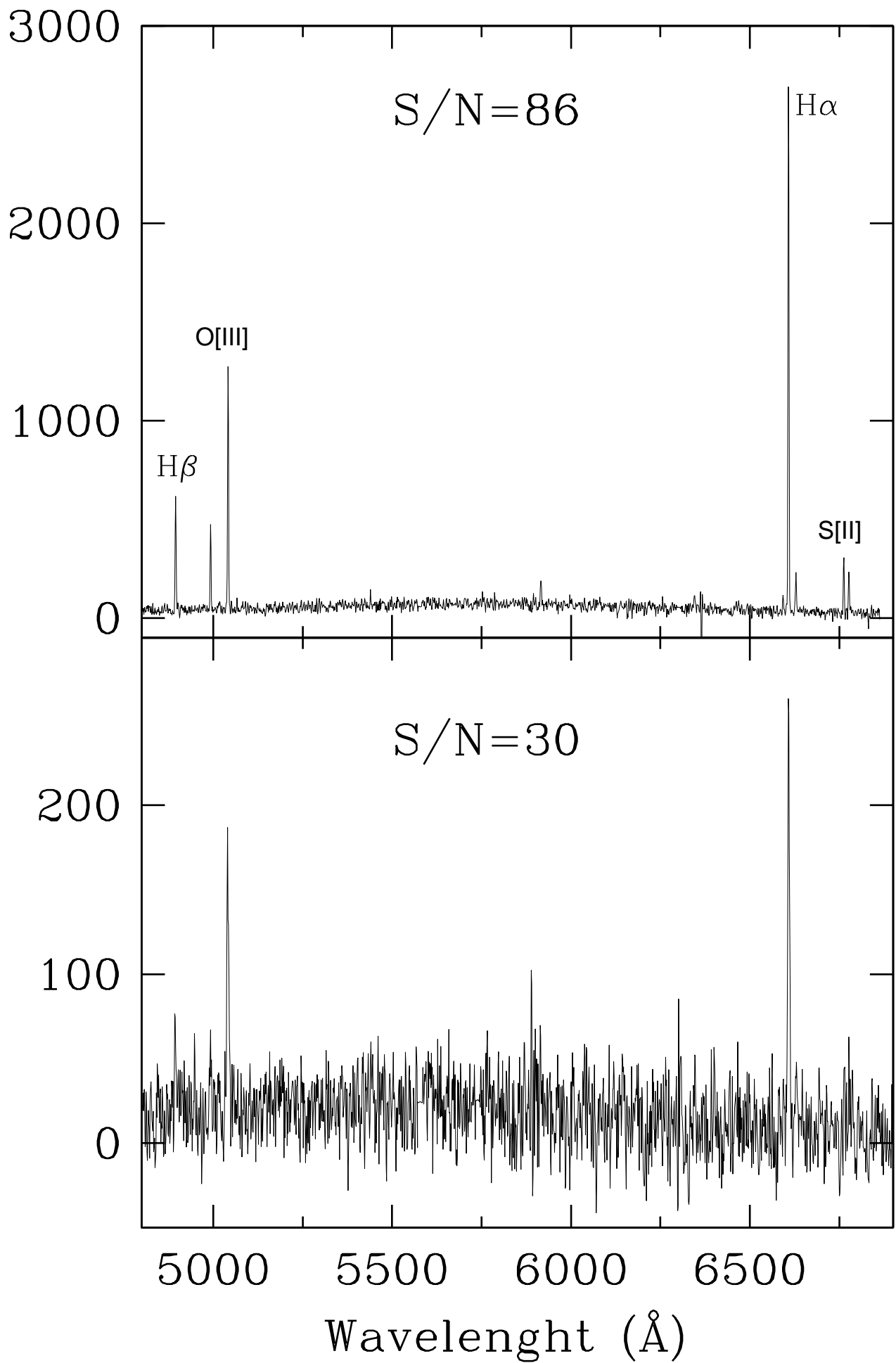
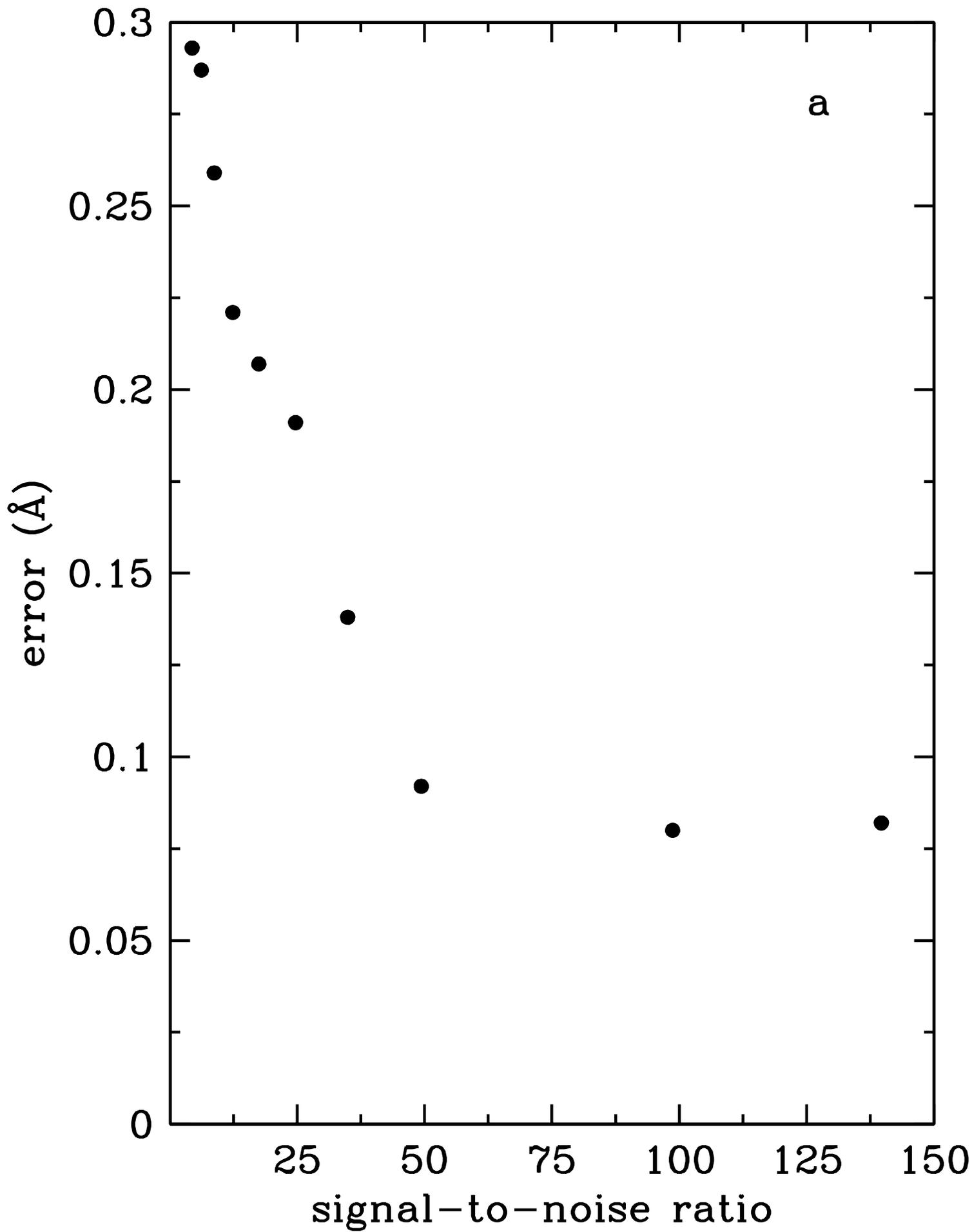
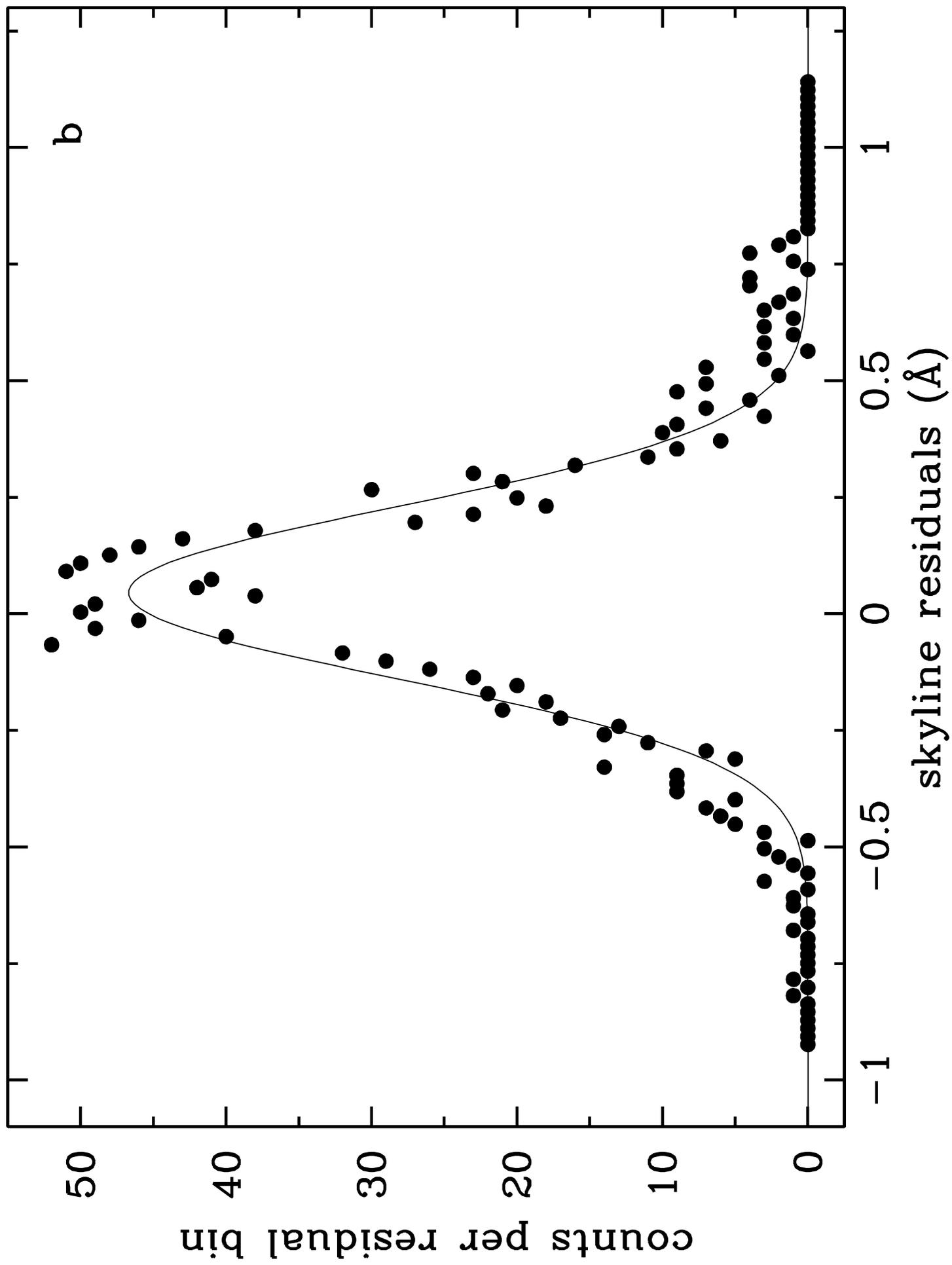
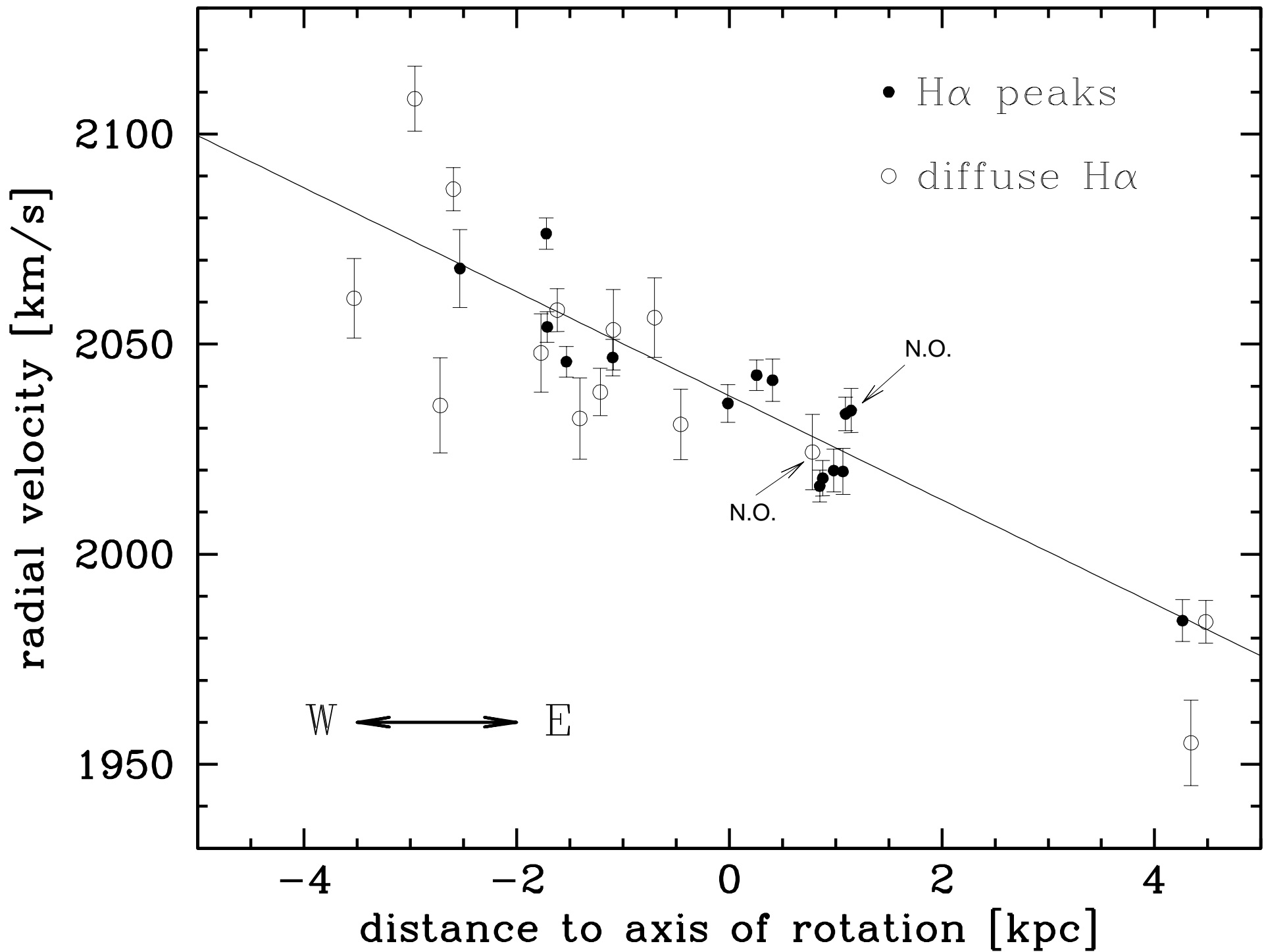
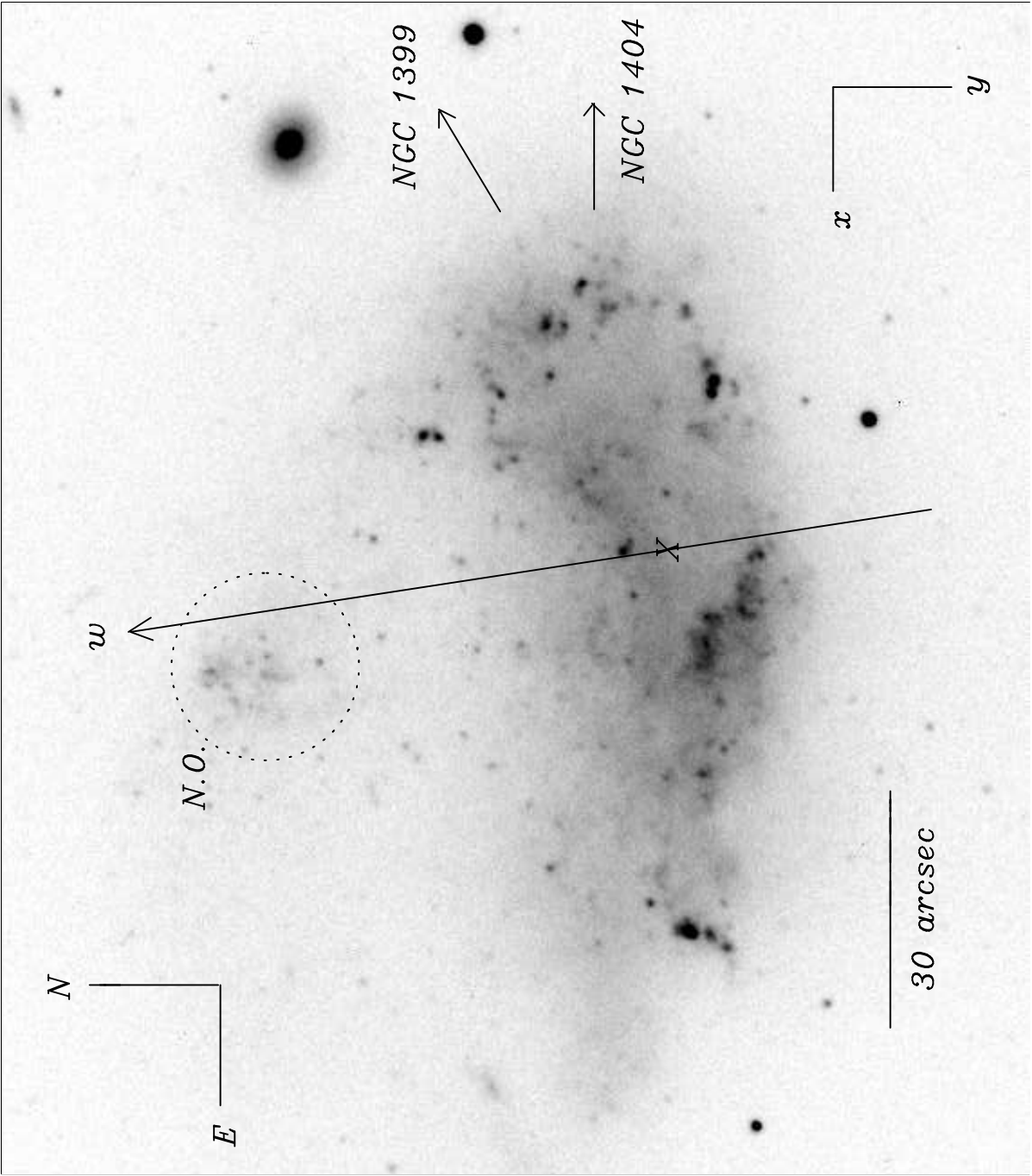
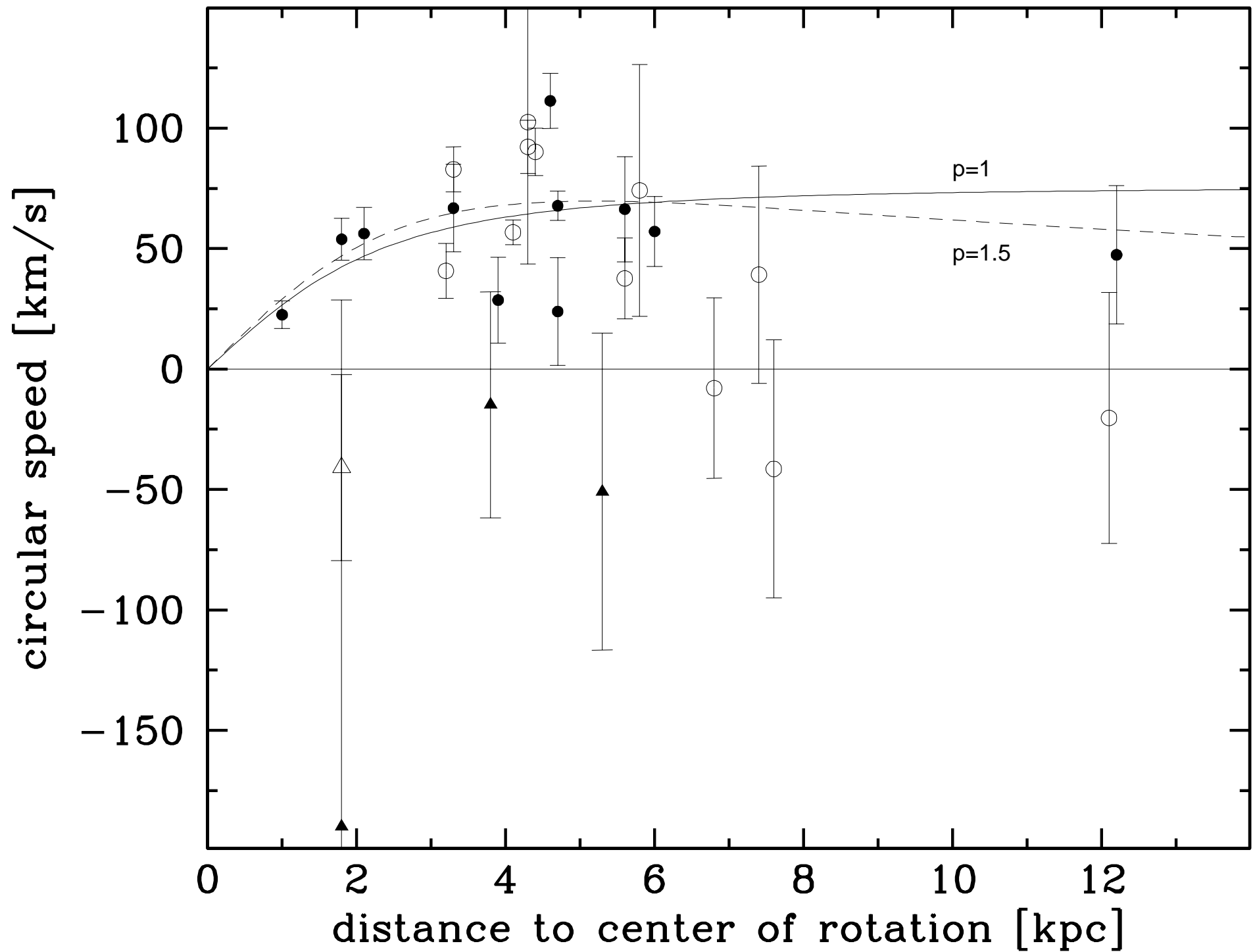
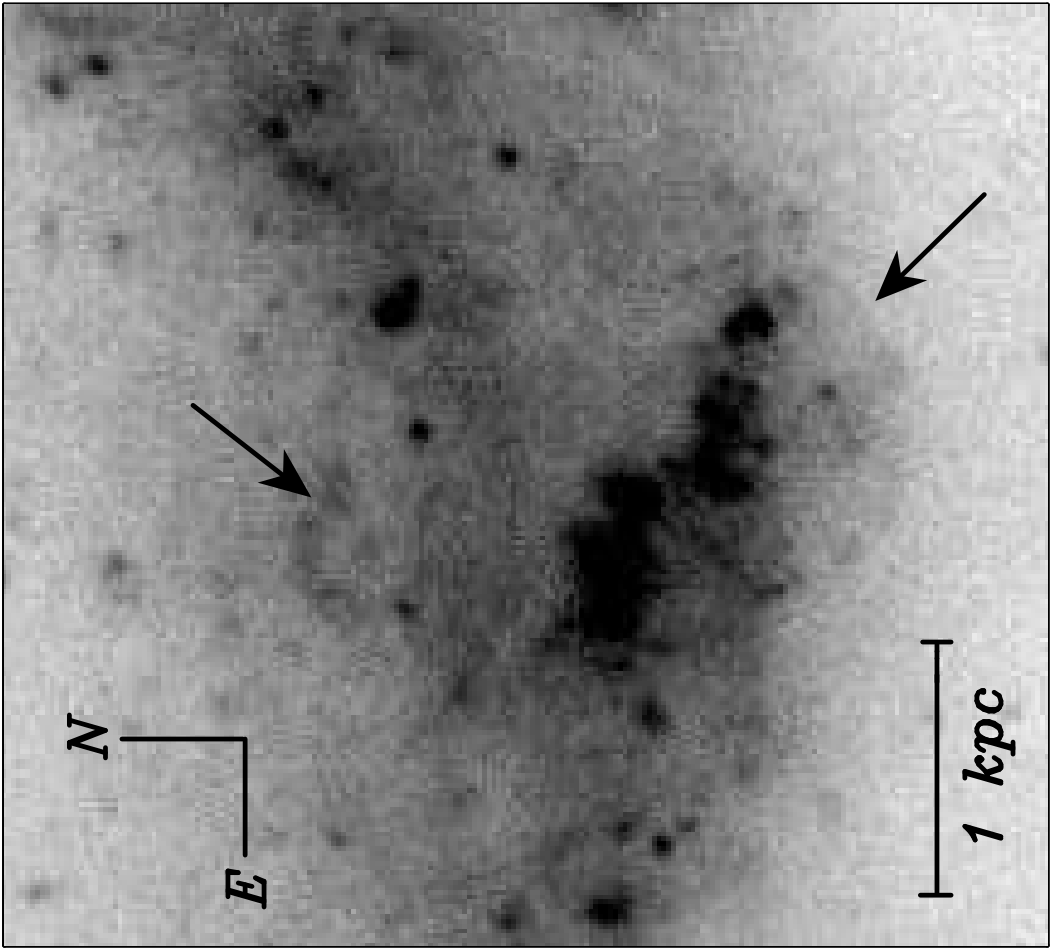
\end{document}